\documentclass[letterpaper]{article} % DO NOT CHANGE THIS
\usepackage{{aaai2026}}
\usepackage{times}  % DO NOT CHANGE THIS
\usepackage{helvet}  % DO NOT CHANGE THIS
\usepackage{courier}  % DO NOT CHANGE THIS
\usepackage[hyphens]{url}  % DO NOT CHANGE THIS
\usepackage{graphicx} % DO NOT CHANGE THIS
\usepackage{natbib}  % DO NOT CHANGE THIS AND DO NOT ADD ANY OPTIONS TO IT
\usepackage{caption} % DO NOT CHANGE THIS AND DO NOT ADD ANY OPTIONS TO IT
\usepackage{algorithm}
\usepackage{algorithmic}
\usepackage{pifont}
\usepackage{booktabs}
\usepackage{arydshln}
\usepackage{subcaption} 
\usepackage{enumitem}
\usepackage{amsmath}
\usepackage{multirow}

\usepackage{listings}
\usepackage{xcolor}

\usepackage{newfloat}
\usepackage{listings}
\DeclareCaptionStyle{ruled}{labelfont=normalfont,labelsep=colon,strut=off} % DO NOT CHANGE THIS
\floatstyle{ruled}
\newfloat{listing}{tb}{lst}{}
\floatname{listing}{Listing}
\title{DualSpeechLM: Towards Unified Speech Understanding and Generation \\via Dual Speech Token Modeling with Large Language Models}
\author{
    Yuanyuan Wang\textsuperscript{\rm 1},
    Dongchao Yang\textsuperscript{\rm 1},
    Yiwen Shao\textsuperscript{\rm 2},
    Hangting Chen\textsuperscript{\rm 2},
    Jiankun Zhao\textsuperscript{\rm 1},\\
    Zhiyong Wu\textsuperscript{\rm 1,3,*},
    Helen Meng\textsuperscript{\rm 1},
    Xixin Wu\textsuperscript{\rm 1}\thanks{Corresponding authors.}
}
\affiliations {
    \textsuperscript{\rm 1}The Chinese University of Hong Kong,
    \textsuperscript{\rm 2}Tencent AI Lab,
    \textsuperscript{\rm 3}Tsinghua University
}
\usepackage{bibentry}
\begin{document}

\maketitle

\begin{abstract}
Extending pre-trained text Large Language Models (LLMs)'s speech understanding or generation abilities by introducing various effective speech tokens has attracted great attention in the speech research community.
However, building a unified speech understanding and generation model still faces the following challenges:
(1) Due to the huge modality gap between speech and text tokens, extending text LLMs to unified speech LLMs relies on large-scale paired data for fine-tuning, and 
(2) Generation and understanding tasks prefer information at different levels, e.g., generation benefits from detailed acoustic features, while understanding favors high-level semantics. This divergence leads to difficult performance optimization in one unified model. 
To solve these challenges, in this paper, we present two key insights in speech tokenization and speech language modeling. Specifically, we first propose an Understanding-driven Speech Tokenizer (USTokenizer), which extracts high-level semantic information essential for accomplishing understanding tasks using text LLMs.
% while simultaneously preserving the ability to recover detailed characteristics for generation tasks by a reconstruction loss. 
In this way, USToken enjoys better modality commonality with text, which reduces the difficulty of modality alignment in adapting text LLMs to speech LLMs.
Secondly, we present DualSpeechLM, a dual-token modeling framework that concurrently models USToken as input and acoustic token as output within a unified, end-to-end framework, seamlessly integrating speech understanding and generation capabilities.
Furthermore, we propose a novel semantic supervision loss and a Chain-of-Condition (CoC) strategy to stabilize model training and enhance speech generation performance.
% by generating tokens with semantic information to serve as conditional steps for the subsequent generation of tokens with detailed acoustic information. 
%generation ensure training stability and improve the model’s capability. A novel Chain-of-Condition (CoC) strategy is further proposed to enhance speech generation performance
% mitigating the gap between speech understanding and generation tasks.
Experimental results demonstrate that our proposed approach effectively fosters a complementary relationship between understanding and generation tasks, highlighting the promising strategy of mutually enhancing both tasks in one unified model.\footnote{
Code and demo: https://github.com/lavendery/UUG.}

\end{abstract}


\section{Introduction}

Recent advancements in autoregressive large language models (LLMs) have demonstrated excellent performance in the natural language processing community \cite{achiam2023gpt, touvron2023llama, team2023gemini}.
Leveraging the powerful foundations of text LLMs, recent advancements have led to emergence of speech LLMs that possess speech understanding and generation capabilities. 
In addition to fine-tuning textual LLMs to separately perform speech understanding tasks~\cite{gong2023joint, tang2023salmonn, chu2023qwenaudio, wang2024blsp, wang2025unisep} and generation tasks~\cite{wang2023neural, anastassiou2024seed, kim2024clam, wang2024speechx, yang2024uniaudio, yang2025simplespeech, jia2025ditar}, developing unified models that excel at both capabilities has been explored in recent years~\cite{zhang2023speechgpt, xie2024miniomni2, fu2024vita, nguyen2025spirit, defossez2024moshi, xu2025qwen2}. However, several limitations still remain.

\begin{figure}[!t]
    \centering
    \begin{subfigure}{0.45\textwidth}
        \centering
        \includegraphics[width=\textwidth]{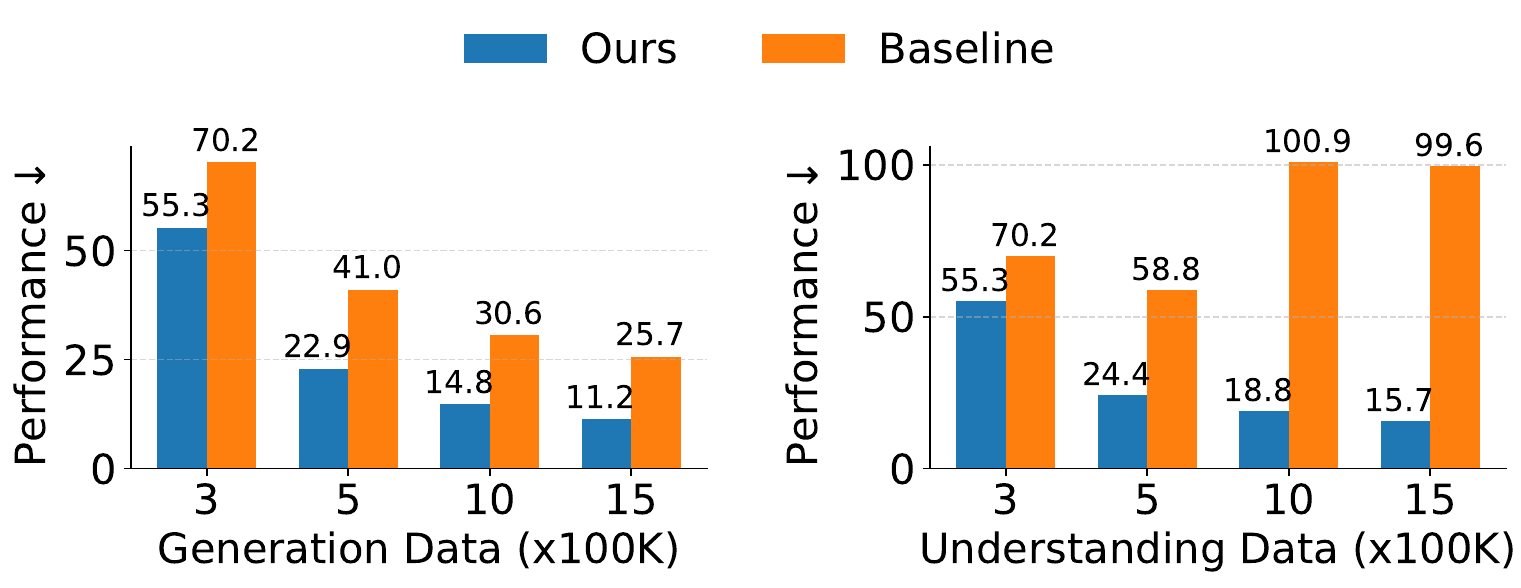}
        \caption{Generation performance ($\downarrow$), measured by generated speech WER (\%), using different amounts of generation (left) or understanding (right) data, with the amount of understanding (left) or generation (right) data fixed as 300K samples.}
        \label{fig:preliminary_tts}
    \end{subfigure}
    % \vspace{0.5em}
    \begin{subfigure}{0.45\textwidth}
        \centering
        \includegraphics[width=\textwidth]{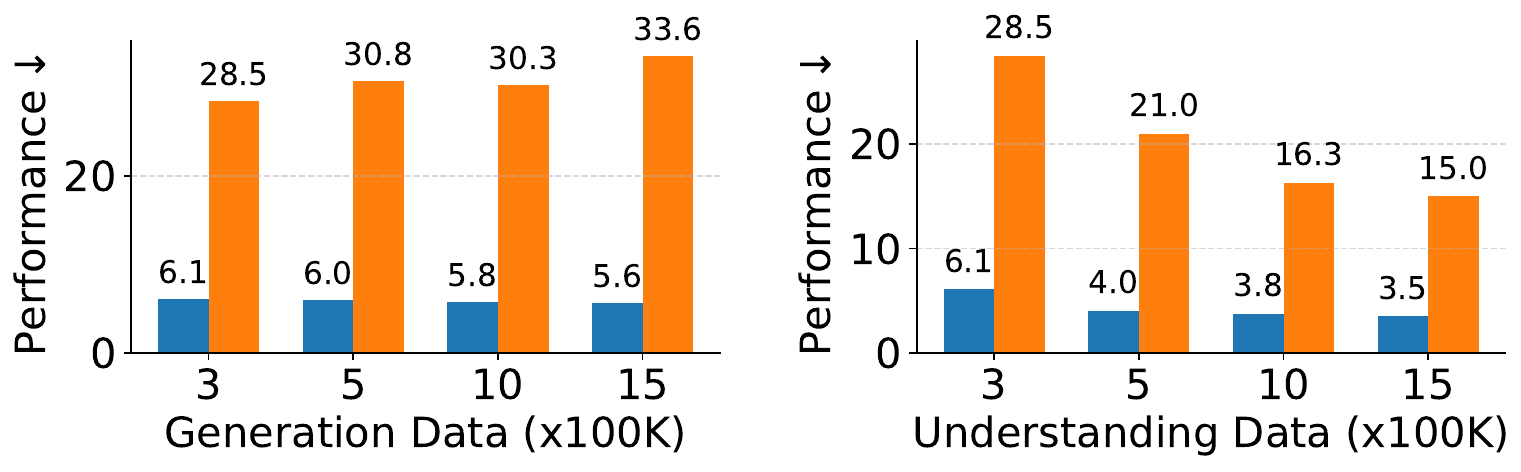}
        \caption{Understanding performance ($\downarrow$), measured by speech recognition WER (\%), using different amounts of generation (left) or understanding (right) data, with the amount of understanding (left) or generation (right) data fixed as 300K samples.}
        \label{fig:preliminary_asr}
    \end{subfigure}
    \caption{Comparison of baseline and our model on generation and understanding tasks with different ratios of generation and understanding training data.}
    \vspace{-10pt}
    \label{fig:preliminary}
\end{figure}

First, adapting pre-trained text LLMs to unified speech LLMs still relies heavily on large-scale paired speech-text data~\cite{zhang2023speechgpt,defossez2024moshi,xie2024miniomni2,xu2025qwen2}.
For example, SpeechGPT~\cite{zhang2023speechgpt} and SpiritLM~\cite{nguyen2025spirit} require approximately 70K and 570K hours of paired data, respectively.
This dependence stems from the substantial modality gap between speech and text, which hinders capability transfer.
Second, existing speech LLMs struggle to meet the distinct informational needs of understanding and generation.
As shown in Figure~\ref{fig:preliminary} (a) Left and (b) Left, the Baseline model—trained solely with acoustic tokens—exhibits a contradiction between tasks, i.e., improving one often leads to degradation of the other, highlighting its inability to balance both tasks effectively.
% Specially, speech signals encompass semantic information (e.g., \textit{intention information}) and diverse additional elements (e.g., \textit{paralinguistic cues}).
In fact, generation tasks demand rich acoustic details (e.g., prosody, emotion, speaker traits) for high-fidelity synthesis~\cite{zeghidour2021soundstream, defossez2022high, kumar2023high, wang2023neural}, which acoustic tokens capture well but lack high-level semantics~\cite{shi2024espnet, dhawan2024codec, anastassiou2024seed, 10447929}.
% making them suboptimal for understanding tasks.
Conversely, understanding tasks benefit from semantic features~\cite{borsos2023audiolm, rubenstein2023audiopalm, maiti2024voxtlm}, but semantic tokens inevitably compromise the acoustic details needed for natural speech generation.

% 目标：
To solve these problems, we propose two key insights from the perspectives of speech tokenization and language modeling.
First, we present an Understanding-driven Speech Tokenizer (USTokenizer) that can extract high-level semantic features, which are critical for understanding tasks.
% while preserving the ability to retrieve fine-grained details necessary for generation tasks. 
Unlike prior methods that rely on only self-supervised learning (SSL) representation quantization~\cite{hsu2021hubert, chen2022wavlm} or automatic speech recognition (ASR)-based objectives~\cite{du2024cosyvoice, zeng2024glm4} to capture semantics, our approach directly aligns speech tokenizer with the semantic understanding capabilities of text LLMs.
This leads to USTokens that have inherently better alignment with text modality, significantly easing modality alignment when adapting text LLMs to speech LLMs.
Secondly, building on USTokenizer, we introduce DualSpeechLM, a novel dual speech token modeling framework that effectively handles the distinct informational needs of understanding and generation within a unified end-to-end architecture.
Unlike conventional approaches that use the same token as input and output of LLM, our model separates them by using USTokens as input and acoustic tokens as output.
Specifically, the USTokenizer provides high-level semantic information to boost understanding tasks, while an AcousticGPT module restores fine-grained acoustic details, enabling diverse and realistic speech generation within a unified end-to-end framework.
In addition, we introduce a semantic supervision loss and a Chain of Condition (CoC) strategy to stabilize training and further improve the performance of the unified framework.
% CoC by promoting the latent consistency space of text and USTokenizer.
In summary, our contributions are as follows:

\begin{itemize}[itemsep=0pt,topsep=0pt,parsep=0pt,leftmargin=10pt]
\item 
We present a USTokenizer that can extract high-level semantic information and reduce the modality gap when adapting text LLMs to speech LLMs.

\item 
We propose an end-to-end dual token modeling framework, DualSpeechLM, that simultaneously accepts USTokens as input and generates acoustic tokens as output, effectively accommodating distinct informational needs.
\item 
We also propose a novel semantic supervision loss and a CoC strategy to improve training stability and generation performance.
\item 
Experiments demonstrate that our method achieves faster convergence and excellent performance with small-scale data, 
% effectively bridging the modality gap, 
and enhances mutual improvements between understanding and generation tasks.
\end{itemize}

\section{Related Work}
\subsection{Speech Tokenization}
The success of autoregressive language models~\cite{achiam2023gpt, touvron2023llama, team2023gemini} has spurred progress in speech LLMs~\cite{chu2023qwenaudio, wang2024speechx, wang2024freeze, defossez2024moshi}, where speech tokenizers are essential for converting continuous signals into discrete tokens~\cite{yangalmtokenizer}.
Speech tokenizers are typically categorized as acoustic or semantic \cite{borsos2023audiolm, parkerscaling, yanguniaudio15}.
Acoustic tokens, optimized for signal reconstruction~\cite{zeghidour2021soundstream, defossez2022high, yang2023hifi, wang2023neural}, capture detailed acoustic features beneficial for generation, but perform poorly on understanding tasks like ASR~\cite{shi2024espnet, dhawan2024codec, anastassiou2024seed, 10447929}. 
Previous semantic tokenizers were trained in two ways: (1) applying clustering \cite{borsos2023audiolm, zhang2023speechgpt, shi2023multi} or vector quantization (VQ) \cite{huang2024repcodec} to representations of SSL models ~\cite{hsu2021hubert, chen2022wavlm}, and (2) applying a VQ layer to the intermediate layer of ASR models ~\cite{du2024cosyvoice, zeng2024glm4, du2024cosyvoice2}. Although these semantic tokenizers have shown benefits for understanding tasks~\cite{borsos2023audiolm, rubenstein2023audiopalm, maiti2024voxtlm}, they do not explicitly consider the alignment to the modality of text LLM~\cite{li2025baichuanaudio}. Furthermore, multi-codebook designs~\cite{zhang2023speechtokenizer, defossez2024moshi, yangalmtokenizer} aim to capture semantics and acoustics in different codebooks jointly, but often introduce complexity when integrated with LLMs. In this work, we present a single VQ-codebook USTokenizer, which not only incorporates high-level semantic information but also achieves better alignment between speech and text modality when applied to LLMs.

\subsection{Speech Language Models}

Recent advances in speech LLMs have explored unified understanding and generation~\cite{zhang2023speechgpt, pan2024auto, defossez2024moshi, nguyen2025spirit}.
Some speech LLMs~\cite{yang2024uniaudio, shi2025balancing} adopt acoustic tokens to ensure high-fidelity speech synthesis. However, such tokens usually lead to degraded performance in understanding tasks, particularly in low-resource scenarios.
By contrast, semantic tokens perform better in understanding tasks. Yet, their lack of acoustic details often results in reduced generation quality.
To compensate for generation, existing works~\cite{polyak2021speech, zhang2023speechgpt, du2024cosyvoice, nguyen2025spirit} introduce additional components such as diffusion models~\cite{ho2020denoising} or flow
matching~\cite{lipman2022flow}, to convert speech tokens
into a Mel spectrogram, and then a HiFi-GAN~\cite{kong2020hifi-gan} vocoder is used to synthesize waveform with the Mel spectrogram as input.
These multi-stage pipelines increase complexity and risk error accumulation~\cite{jia2025ditar}.
To address these limitations, we propose DualSpeechLM, an end-to-end dual-token modeling framework that explicitly models USTokens as input for understanding and acoustic tokens as output for generation, effectively achieving distinct informational requirements.
\section{Methodologies}
\iffalse
This section presents the technical details of the proposed method, 
% As shown in Figures \ref{fig:tokenizer} and \ref{fig:overall}, the model architectures of USTokenizer and DualSpeechLM are illustrated. 
with model overviews of USTokenizer and DualSpeechLM shown in Figures~\ref{fig:tokenizer} and \ref{fig:overall}.
% Next, we will describe the architecture of our USTokenizer and DualSpeechLM, respectively.
% This section introduces the proposed method. 
% We describe each component in detail below.
\fi

We propose a novel unified speech LLM framework, comprising an understanding-driven speech tokenization method and a dual-token modeling paradigm. This framework enhances both understanding and generation capabilities in the resulting DualSpeechLLM. In the following, we describe each module of our framework.

\subsection{USTokenizer}\label{sec_dualtokenizer}
% Previous studies \cite{borsos2023audiolm, rubenstein2023audiopalm} have demonstrated that semantic tokens are more easily modeled by text-based LLMs due to their incorporation of richer linguistic information. However, previous approaches \cite{zhang2023speechgpt, du2024cosyvoice, zeng2024glm4} typically overlook the need for more low-level detail information, which is vital for generation tasks. Moreover, they focus primarily on optimizing the tokenizer model itself, without considering whether these tokens are suitable for training downstream tasks in LLMs.

\begin{figure}[htb]
\centering
\includegraphics[width=8.3cm]
{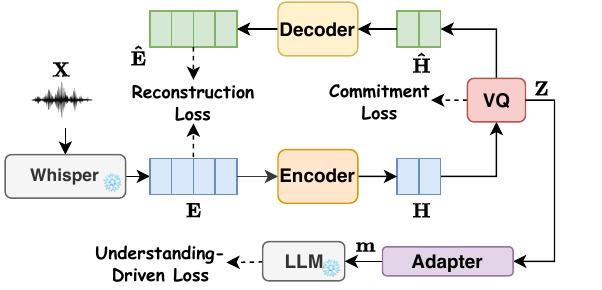}
\caption{The architecture of USTokenizer, which can extract high-level semantic features aligned with text LLMs via an understanding-driven loss.}
\label{fig:tokenizer}
\end{figure}

As shown in Figure \ref{fig:tokenizer}, our Understanding-driven Speech Tokenizer (USTokenizer) consists of a pre-trained Whisper encoder, another downsampling Encoder, a vector quantizer (VQ), an upsampling Decoder, an Adapter module, and a frozen text LLM.
% Following previous semantic token extraction methods~\cite{zhang2023speechgpt,du2024cosyvoice,zeng2024glm4}, 
Given a speech utterance $\mathbf{X}$, we use the final hidden states $\mathbf{E}$ from the Whisper-medium encoder~\footnote{https://github.com/openai/whisper} as the input to the downsampling Encoder and then the VQ to obtain the Understanding-driven Speech Token (USToken).
% 2$\times$ downsampling convolutional layer
Additionally, we novelly project the quantized vector from VQ to the input space of LLM via an adapter to integrate the guidance provided by text LLMs to optimize the speech token during training, as described in the following sections.
\iffalse
which comprises linear and convolution layers, for training on a series of understanding tasks.
% including ASR, Speech Emotion Recognition (SER), and Speech Question Answering(SQA). 
This strategy ensures that the speech token extracted by VQ is more closely aligned with the feature space of text LLMs, 
% enhancing modality commonality with text tokens, 
significantly reducing the alignment complexity when adapting text LLMs to unified speech LLMs.
% making it better suited for subsequent large-scale downstream tasks and improving performance.
\fi

% \subsubsection{Encoder}
% The Encoder module in our DualTokenizer is designed to process the downsampling representations $D$ and extract meaningful features $H$. 
% It consists of two convolutional blocks, each designed to capture hierarchical representations of the input embeddings through two residual units and convolutions. 
% The output $H$ from the last convolutional block is then passed to the VQ module for discretization.
% \subsubsection{Decoder}
% The Decoder module in DualTokenizer is designed to process the vector $\mathbf{Z}$ and recover them into the original feature space $\hat{D}$. Symmetrical with the Encoder, it consists of two convolutional blocks. 
% Each block contains a convolution followed by residual units with dilated convolutions. 
% The final output of the blocks is passed through a convolutional layer to produce the feature map $\hat{D}$.

\subsubsection{Encoder and Decoder}
The Encoder converts Whisper features $\mathbf{E}$ into representations $\mathbf{H}$ using a 2× downsampling convolution followed by two residual convolutional blocks, 
% These representations are then discretized by the VQ module.
then discretized by VQ.
Symmetrically, the Decoder reconstructs $\mathbf{\hat{E}}$ from quantized vectors $\mathbf{\hat{H}}$.
Both Encoder and Decoder adopt similar residual convolutional blocks with RepCodec~\cite{huang2024repcodec}.
%Inspired by RepCodec~\cite{huang2024repcodec}, to retain fine-grained semantic details, the quantized vectors $\mathbf{\hat{H}}$ are passed through a Decoder to perform 2$\times$ upsampling and reconstruct the Whisper encoder outputs as $\hat{E}$.
We then compute the reconstruction loss via mean squared error (MSE) between $\mathbf{E}$ and $\mathbf{\hat{E}}$, formulated as:
\begin{equation}\label{eq_tokenizer_recon}
\begin{aligned}
\mathcal{L}_{\text{reconstruction}} (\mathbf{E}, \mathbf{\hat{E}}) = \frac{1}{N} \sum_{i=1}^{N} (E_i - \hat{E_i})^2
\end{aligned},
\end{equation}
where $N$ is the number of elements in the embedding vector.

\begin{figure*}[!t]
\centering
\includegraphics[scale=0.85]
{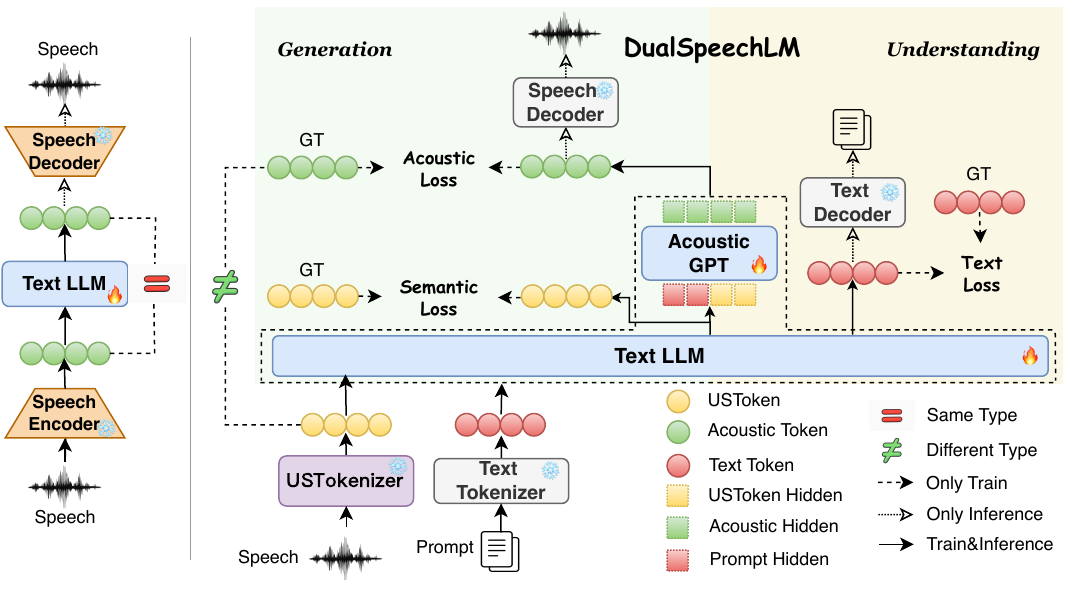}
\vspace{-10pt}
\caption{
% The left part illustrates the framework of the baseline, while the right part provides an overview of the proposed DualSpeechLM.
DualSpeechLM’s dual-token modeling paradigm. The left illustrates the baseline pipeline treating LLM input/output as identical tokens. 
In contrast, our DualSpeechLM (right) incorporates an Acoustic GPT module into the text LLM module for joint training, separately processing USToken inputs and acoustic token outputs through distinct modeling paths, effectively capturing the different levels of information required for both generation and understanding tasks.
}
\label{fig:overall}
\vspace{-8pt}
\end{figure*}

\begin{table*}[tb]
\small
\centering
\setlength{\tabcolsep}{1.1mm}
\begin{tabular}{cccccccccc}
\toprule
\multirow{3}{*}{\textbf{Model}}            & \multirow{3}{*}{\textbf{LLM}}  
& \multirow{3}{*}{\textbf{Input Token}}   
& \multirow{3}{*}{\textbf{Training}}
& \multirow{3}{*}{\textbf{Data(hrs)}}
&  \multicolumn{2}{c}{\textbf{ASR}}                                       & \textbf{S2TT}                                      & \textbf{SER}               & \textbf{SQA}   \\ 
\cline{6-10} 
&  &   &      &                     & Clean             & Other             & \multicolumn{1}{c}{En2De}                                   & -                 & -                        \\ 
                                  % \cline{4-9} 
                               &   &                          &             &              & \multicolumn{2}{c}{$w\downarrow$}                       & $b4\uparrow$                                  & $acc\uparrow$                & $b4\uparrow/gs\uparrow$        
\\ 
                                  \midrule

% \textbf{Unified Model} \\

SpeechGPT
~\cite{zhang2023speechgpt}
& LLaMA-7B
% ~\cite{touvron2023llama} 
&  D(Hubert)
% ~\cite{hsu2021hubert} 
& Full FT 
% $\rightarrow$ LoRA
&            70K &                            \multicolumn{1}{c}{42.73}                  & 78.54                  & \multicolumn{1}{c}{1.07}                                        & -                   & 3.58/40  \\

SpiritLM
~\cite{nguyen2025spirit}
& LLaMA-7B 
&  D(Hubert)
& Full FT
& 570K
&                                        \multicolumn{1}{c}{6.0}                  & 11.0                                   & -                   & -                   & -                  \\

Mini-Omni2
~\cite{xie2024miniomni2}
& Qwen2-0.5B 
&  C(Whisper)
& Full FT
&           9K &                             4.8                  & 9.8                  & -                                        & -                   & -                   \\

VITA
~\cite{fu2024vita}
& Mixtral-8x7B 
&  C(Mel)
& Full FT
&      20K    &                              \multicolumn{1}{c}{8.14}                  & 18.41                  & -                    & -                    & 7.62/70          \\

Moshi
~\cite{defossez2024moshi} 
& Helium-7B & 
D(Mimicodec)
& Full FT
& 7M & 5.7 & - & - & - & -
\\

Qwen2.5-Omni
~\cite{xu2025qwen2}
& Qwen2.5-7B
& C(Whisper) 
& Full FT
&$\star$
& 1.8 & 3.4
& 30.2 & 60.03 & 4.28/60

\\

\hline

Baseline-Acoustic & Phi3.5-3B                                       &  D(Wavtokenizer) 
& LoRA 
&  4.5K &                                      36.52                  & 80.06                 & 1.91                                       & \underline{54.90}                   & 17.68/76                  \\ 

Baseline-Semantic & Phi3.5-3B                                       &  D(Hubert) 
& LoRA 
&  4.5K &    5.70 &  \underline{14.32} & \underline{11.13} & 51.91 & 42.01/85                                                  \\ 
\hdashline

\multirow{2}{*}{\textbf{DualSpeechLM}}            & \multicolumn{1}{c}{\multirow{2}{*}{Phi3.5-3B}} & \multirow{1}{*}{D(Hubert)} 
& LoRA 
& 4.5K
& \underline{5.56}                   & 14.62                  & \multicolumn{1}{c}{10.20}                                       & 51.77                   & \underline{42.59}/\underline{85}                   \\   
& \multicolumn{1}{c}{}                                       & \multirow{1}{*}{\textbf{D(USToken)}}  
     & LoRA 
     &4.5K & \textbf{4.22} & \textbf{9.71}                   & \textbf{19.74} & \textbf{60.92}                   & \textbf{44.38/88} \\ 
                                \bottomrule
\end{tabular}
\caption{
% The understanding performance comparison. 
% `D' denotes Discrete and `C' denotes Continuous. $w, b4, gs$ represent WER, BLEU4, ChatGPTScore, respectively. 
% `Full FT' denotes full fine-tuning, where all model parameters are updated, whereas `LoRA' (Low-Rank Adaptation) represents a parameter-efficient approach that updates only minimal trainable parameters.
% $^\star$Since BLSP does not support SER tasks, we instead use the results from the single-task model BLSP-Emo~\cite{wang2024blspemo}.
% $\star$Qwen2.5-Omni uses 300 billion audio tokens, while our DualSpeechLM utilizes only 405 million tokens.
Understanding capability evaluation.
`D': Discrete, `C': Continuous, `Full FT': full fine-tuning.
$w$, $b4$, $gs$ denote WER, BLEU4, ChatGPTScore.
% $^\star$BLSP does not support SER tasks, we use the results from the single-task model BLSP-Emo~\cite{wang2024blspemo}.
$^\star$Qwen2.5-Omni uses 300B audio tokens, and DualSpeechLM uses 405M speech tokens.}
\label{main_for_understanding}
\end{table*}

\begin{table*}[tb]
\small
\centering
\setlength{\tabcolsep}{0.7mm}
\begin{tabular}{cccccccccc}
\toprule
      \multirow{3}{*}{\textbf{Model}}             &    \multirow{3}{*}{\textbf{LLM}}            &  \multirow{3}{*}{\textbf{Input Token}}  
    & \multirow{3}{*}{\textbf{Training}}
      & \multicolumn{2}{c}{\textbf{TTS}}               & \textbf{VC}              & \multicolumn{2}{c}{\textbf{T2ST}}                                                          & \textbf{SC}        \\ \cline{5-10} 
           &              &            &                           & \multicolumn{1}{c}{Clean}           & Other           & VCTK            & \multicolumn{1}{c}{Es2En} & Fr2En & - \\ 

% \cline{4-9} 
                        & &  &     & \multicolumn{2}{c}{$s\uparrow/w\downarrow/d\uparrow$}
                                  & $s\uparrow/w\downarrow/d\uparrow$
                                  & \multicolumn{2}{c}{$b4\uparrow$}
                                  & $b4\uparrow/gs\uparrow$
                                  \\
           
           \midrule
GT                        &        & \multicolumn{1}{c}{-}                         & \multicolumn{1}{c}{-}                                                                            & \multicolumn{1}{c}{1.0/3.94/3.86}   & 1.0/5.35/3.78   & 1.0/2.64/3.62   & \multicolumn{1}{c}{-}                             & -                             & -/-        \\ 

\hline
% \textbf{Unified Model} \\
\multicolumn{1}{c}{SpeechGPT}
% ~\cite{zhang2023speechgpt}
& LLaMA-7B
% ~\cite{touvron2023llama} 
&  D(Hubert)
% ~\cite{hsu2021hubert} 
& Full FT
 & \multicolumn{1}{c}{-/22.15/3.97} & -/24.55/3.96 & - & 14.62                          & 14.74                          & 3.50/54      \\

\multicolumn{1}{c}{Mini-Omni2}
% ~\cite{xie2024miniomni2}}
& Qwen2-0.5B 
&  C(Whisper)
& Full FT
& - & - & -& -                          & -                          & 2.22/54      \\

% \multicolumn{1}{c}{ SpiritLM}
% % ~\cite{nguyen2025spirit}
% & LLaMA2-7B 
% &  Hubert
% & \multicolumn{1}{c}{-} & - & - & -                          & -                          & -        \\

Moshi
& Helium-7B & D(Mimicodec)
& Full FT
& - & - & -& -                          & -                          & 1.75/50 
\\

Qwen2.5-Omni
% ~\cite{xu2025qwen2}
& Qwen2.5-7B
& C(Whisper) 
& Full FT
& -/3.73/4.10$^\star$& -/4.29/4.08 &-&-&-& 6.70/70

\\
\hline

Baseline-Acoustic & Phi3.5-3B                                        & D(Wavtokenizer)    
& LoRA & \underline{0.88}/22.11/\underline{3.76} & \underline{0.87}/26.38/\underline{3.69} & 0.80/22.07/3.30 & 8.52                          & 7.62                         & 0.52/62       \\ 

Baseline-Semantic & Phi3.5-3B                                        & D(Hubert)    
& LoRA &  0.80/21.72/3.29 & 0.81/22.32/3.26 & \underline{0.81}/18.88/3.25 & \underline{18.05} & 15.76 & 10.44/60
\\ 
\hdashline
  \multirow{2}{*}{\textbf{DualSpeechLM}}                                & \multirow{2}{*}{Phi3.5-3B}                          & D(Hubert)    &LoRA                                                                       & \multicolumn{1}{c}{0.81/\underline{20.80}/3.35} & 0.81/\underline{17.12}/3.38 & \textbf{0.82}/\underline{18.70}/\underline{3.37} & 17.54                        & \underline{15.97}                         & \underline{11.06}/\underline{65}       \\  
                                 
                                  & \multicolumn{1}{c}{}                          &  \textbf{D(USToken)}    
   &LoRA                  & 
                    \textbf{0.90}/\textbf{9.25}/\textbf{3.86}  & \textbf{0.88}/\textbf{9.88}/\textbf{3.82}  & 0.80/\textbf{10.16}/\textbf{3.46} & \multicolumn{1}{c}{\textbf{26.77}}                         & \textbf{23.82}                         & \textbf{16.24}/\textbf{67}      \\  \bottomrule

\end{tabular}
\caption{
% The generation performance comparison. \textbf{Data(hrs)} follows the same setting as Table~\ref{main_for_understanding}. $s, w, d, b4, gs$ denote SIM, WER, DNSMOS, BLEU4, ChatGPTScore respectively.
% $^\star$Qwen2.5-Omni does not support standalone TTS task inference, so we use the API of Qwen-TTS (\url{https://qwenlm.github.io/blog/qwen-tts/}) to infer.
Generation capability evaluation. Data usage is the same as \textbf{Data(hrs)} in Table~\ref{main_for_understanding}.
$s$, $w$, $d$, $b4$, $gs$ denote SIM, WER, DNSMOS, BLEU4, ChatGPTScore.
$^\star$Qwen2.5-Omni lacks standalone TTS inference, so the Qwen-TTS API is used instead.
}
\label{main_for_generation}
\end{table*}

\begin{table}[tb]
\small
\centering
\setlength{\tabcolsep}{0.4mm}
\begin{tabular}{ccc}
\toprule
      \textbf{Model}  
      & \textbf{TTS} ($Q\uparrow/S\uparrow$)              & \textbf{VC} ($Q\uparrow/S\uparrow$)     \\
      % \cline{3-5} 
      %      &       &     \multicolumn{1}{c}{Clean}           & Other           & VCTK           \\ 

           \midrule
GT &     $4.20_{\pm0.33}/3.94_{\pm0.42}$      & $4.48_{\pm0.23}/4.60_{\pm0.25}$        \\ 

\hline
% \textbf{Unified Model} \\
\multicolumn{1}{c}{SpeechGPT}
% ~\cite{zhang2023speechgpt}
&     $3.53_{\pm0.51}$/-     & -
    \\

Qwen-TTS
% ~\cite{xu2025qwen2}
&     $4.49_{\pm0.21}$/-      & -

\\
\hdashline

Baseline-Acoustic    &     $3.67_{\pm0.48}$/$\mathbf{3.77_{\pm0.52}}$      & $3.63_{\pm0.33}$/$3.43_{\pm0.42}$   \\ 

Baseline-Semantic &     $2.26_{\pm0.53}$/$2.76_{\pm0.45}$      & $2.75_{\pm0.0.56}$/$2.93_{\pm0.57}$  
\\ 

  \textbf{Ours-Hubert} &     $3.48_{\pm0.61}$/$3.59_{\pm0.47}$      & $3.54_{\pm0.34}$/$3.18_{\pm0.36}$                                           \\  
\textbf{Ours-USToken}   &     $\mathbf{3.89_{\pm0.31}}$/$3.74_{\pm0.24}$      & $\mathbf{3.80_{\pm0.28}}$/$\mathbf{3.55_{\pm0.37}}$
  \\  \bottomrule

\end{tabular}
\caption{
Subjective evaluation on generation tasks.
$Q, S$ denote QMOS, SMOS.
% Ours-Hubert means DualSpeechLM with Hubert as input; Ours-USToken means DualSpeechLM with USToken as input token.
Ours-Hubert and Ours-USToken refer to DualSpeechLM using Hubert and USToken as input tokens, respectively.
Configurations are the same as Table~\ref{main_for_generation}.
% $^\star$Qwen2.5-Omni lacks standalone TTS inference, so the Qwen-TTS API is used instead.
}
\label{main_for_generation_subjective}
\end{table}

\subsubsection{Vector Quantization}
The VQ module discretizes continuous feature representations $\mathbf{H}$ by mapping them to a codebook of learned vectors. In this work, we use a single VQ layer to quantize the feature $\mathbf{H}$ into discrete tokens $\mathbf{Z}$. Following previous works \cite{huang2024repcodec}, we add a commitment loss $\mathcal{L}_{\text{commit}}$ to ensure training stability, with more details provided in \textit{Appendix B}.

% In addition, the perplexity is computed to measure the diversity and spread of the quantization choices across the codebook \cite{huang2024repcodec}:
% % Perplexity is a metric that reflects the entropy of the distribution of quantized indices. It is defined as:
% \begin{equation}\label{eq_tokenizer_perp}
% \begin{aligned}
% \mathcal{L}_{\text{perplexity}} = \exp \left( - \sum_{i=1}^{N} p_i \log(p_i) \right)
% \end{aligned}
% \end{equation}
% where $p_i$ is the probability of selecting the $i$-th embedding vector from the codebook, computed as the average frequency of the $i$-th codebook vector being chosen during quantization,
% $N$ is the total number of vectors in the codebook.
% A lower perplexity indicates that the quantizer is using a smaller number of vectors from the codebook, while a higher perplexity suggests a more uniform usage of the codebook vectors.

\subsubsection{Understanding-Driven Loss}
The LLM module aims to align the speech token $\mathbf{Z}$ with the LLM's input space. 
By optimizing the understanding tasks upon the text LLM, the required understanding capability will be backpropagated to the optimization of speech token, such that the modality gap between speech and text tokens can be effectively reduced.
% such as ASR, SER, and SQA.
The LLM-based understanding loss is formulated as the likelihood of generating the target response given a speech prompt using a text LLM (e.g., generating answers given the prompt of speech question):
\begin{equation}\label{eq_tokenizer_under}
\begin{aligned}
\mathcal{L}_{\text{Under}} = -\sum_{t=1}^{L} \log p(S_t\mid\mathbf{m}, S_{<t}; \theta),
\end{aligned}
\end{equation}
where $L$ is the length of the sequence, $S_t$ represents the target token at position $t$, $S_{<t}$ is the sub-sequence of text tokens before $t$,
and $\mathbf{m}$ is the feature of speech prompt extracted by USTokenizer.
$\theta$ represents parameters of the text LLM.
% $p_\theta(S_t \mid \mathbf{m}, S_{<t})$ denotes the probability of token $S_t$ given the preceding tokens.
% This loss trains the entire system end-to-end, ensuring the VQ space aligns with LLM's text space. 

% In summary, the LLM module not only facilitates alignment of VQ space with LLM's text space but also helps ensure that USTokens extracted by USTokenizer are mapped to feature space compatible with LLMs, thereby facilitating improved performance on subsequent unified understanding and generation tasks.

The LLM module aligns the VQ space with text LLM's input space, which means USTokens are mapped to an LLM-compatible feature space, enhancing subsequent unified understanding and generation modeling.
The final total loss of USTokenizer is as follows:
\begin{equation}\label{eq_tokenizer_allloss}
\begin{aligned}
\mathcal{L}_{\text{USTokenizer}} = \alpha \cdot \mathcal{L}_{\text{commit}} + \beta \cdot \mathcal{L}_{\text{Under}} + \gamma \cdot \mathcal{L}_{\text{reconstruction}},
\end{aligned}\nonumber
\end{equation}
where $\alpha, \beta$ and $\gamma$ are weighting hyperparameter.

\subsection{DualSpeechLM}
\label{sec_duallm}

% \subsubsection{Preliminary}
% As shown in the left of Figure \ref{fig:overall}, existing Speech LLMs~\cite{yang2024uniaudio, wang2024speechx, shi2025balancing} typically treat the input and output as the same target token. 
% Methods such as UniAudio \cite{yang2024uniaudio} and SpeechX \cite{wang2024speechx} directly model acoustic tokens. 
% The acoustic tokens contain detailed acoustic information, which leads to poor performance on understanding tasks. 
% There are also two-stage approaches, such as SpeechGPT \cite{zhang2023speechgpt} and SpiritLM \cite{nguyen2025spirit}, which first train a tokenizer as a speech encoder to extract discrete semantic tokens, and then require additional module to restore high-fidelity speech. 
% This complex multistage training and inference process cannot be directly optimized in LLMs, leading to error accumulation. 
% Moreover, the diversity in speech recovery directly sampled from semantic tokens is insufficient.
% \subsubsection{Overview}
% We propose a novel end-to-end solution, DualSpeechLM, which does not require multistage training, 
% as shown in the right part of Figure \ref{fig:overall}. 
% The biggest difference with the left is that DualSpeechLM treats the input and output as distinct targets, modeling USToken as input and acoustic token as output separately by incorporating AcousticGPT into the text LLM module for joint training, thus effectively accommodating different levels of information required for generation and understanding tasks.

As shown in Figure \ref{fig:overall}, unlike prior Speech LLMs that use the same token for both input and output~\cite{yang2024uniaudio, wang2024speechx, shi2025balancing}, DualSpeechLM introduces a novel dual-token design by modeling USTokens as input and acoustic tokens as output via an integrated AcousticGPT, effectively accommodating different levels of information required for understanding and generation tasks.
%Next, we will provide detailed descriptions for DualSpeechLM.

For speech understanding tasks, the USTokenizer first encodes raw speech into USTokens, which are combined with task-specific prompts and fed into the text LLM.
The model is trained using Cross-Entropy (CE) loss between predicted and ground-truth text tokens.
% This ensures that the understanding task aligns with the expected result as defined by the prompt.
During inference, the model generates text tokens conditioned on USTokens and prompts. The text tokens are then decoded into the final outputs.

For the generation task, let 
% $\mathbf{D}$ 
$\mathbf{U^{tar}}$ and $\mathbf{A^{tar}}$ denote USTokens and acoustic tokens of target speech, 
% $\mathbf{C}$
$\mathbf{U^{in}}$
and $\mathbf{P}$ represent the USTokens and prompt of input speech. As shown in Figure~\ref{fig:overall}, the text LLM first predicts $\mathbf{U^{tar}}$ conditioned on $\mathbf{P}$ and $\mathbf{U^{in}}$, formulated as:
\begin{equation}\label{eq_llm_semanticloss1}
\begin{aligned}
p(\mathbf{U^{tar}} | \mathbf{P}, \mathbf{U^{in}}; \theta) = \prod_{t=1}^{L} p(\mathbf{U^{tar}}_t | \mathbf{U^{tar}}_{<t}, \mathbf{P}, \mathbf{U^{in}}; \theta),
\end{aligned}
\end{equation}
where $L$ is the length of $\mathbf{U^{tar}}$, $\mathbf{U^{tar}}_t$ is the target USToken at position $t$ and $\mathbf{U^{tar}}_{<t}$ represents all previous USTokens. $\theta$ denotes model parameters.
To improve training stability and model performance, we introduce a semantic supervision loss:
% Subsequently, we propose a semantic supervision loss formulated as Equation \ref{eq_llm_semanticloss2}. 
% It helps stabilize the training process and improves the performance of both understanding and generation tasks. 
% The specific impact will be detailed in the ablation study section.
\begin{equation}\label{eq_llm_semanticloss2}
\begin{aligned}
\mathcal{L}_{\text{semantic}} = -\frac{1}{L}\sum_{t=1}^{L} \log 
p(\mathbf{U^{tar}}_t | \mathbf{U^{tar}}_{<t}, \mathbf{P}, \mathbf{U^{in}}; \theta)
% p(\mathbf{D}_t | \mathbf{D}_{<t}, \mathbf{P}, \mathbf{C}; \theta).
\end{aligned}
\end{equation}

Next, the predicted USToken sequence $\mathbf{U^{tar}}$ is passed to AcousticGPT, which autoregressively generates acoustic tokens. They are then converted into speech via a Speech Decoder.
For both understanding and generation, we adopt text LLM as backbone and fine-tune it using parameter-efficient Low-Rank Adaptation (LoRA)~\cite{hu2022lora}.
Within the unified framework, we integrate AcousticGPT into text LLM to enable joint end-to-end training. This allows DualSpeechLM to simultaneously capture high-level semantics for understanding and fine-grained acoustic details for generation, which we refer to as dual-token modeling.
% Since both AcousticGPT and Text LLMs are jointly trained in an end-to-end manner, DualSpeechLM can simultaneously focus on the semantic information required for understanding tasks and the acoustic details essential for generation tasks. 
% This strategy is highly versatile, as the acoustic tokens and Speech Decoder can be any open-source codec model. 
Notably, the acoustic tokens and Speech Decoder are modular and can be implemented using any open-source codec model.

\subsubsection{AcousticGPT}
Inspired by AudioLM~\cite{borsos2023audiolm}, we design an AcousticGPT module to autoregressively generate acoustic tokens conditioned on semantic hidden states and speaker embeddings.
% It leverages a GPT-style transformer architecture to predict acoustic details that align with both the semantic content and speaker characteristics.
Built on a GPT-style transformer, the model consists of six causal blocks, each containing a causal self-attention layer followed by a Multi-Layer Perceptron (MLP).
The multi-head self-attention captures dependencies under causal constraints.
% while the MLP refines the embeddings for fine-grained acoustic token generation.
% ensuring causal relationships for the autoregressive token
In addition, each block includes residual connections and layer normalization to enhance training stability and model performance. 
Further details are provided in \textit{Appendix B}.

\subsubsection{Chain of Condition}
To improve training robustness, we introduce a Chain of Condition (CoC) strategy for AcousticGPT.
Specifically, given the dynamically changing semantic hidden states, we randomly sample conditioning signals from three sources with equal probability: prompt hidden states $\mathbf{S}_p$, predicted USTokens $\mathbf{U^{tar}}$, or their concatenation $[\mathbf{S}_p; \mathbf{U^{tar}}]$.
During inference, we consistently use the concatenated form $[\mathbf{S}_p; \mathbf{U^{tar}}]$. 
This CoC strategy acts as a regularizer, preventing overfitting to any single modality and improving the model’s adaptability by exposing it to varied conditioning signals during training. 
Importantly, it also mitigates the adverse effects of inaccurate USToken predictions by allowing the model to flexibly rely on more stable prompt hidden states or concatenation.
As a result, CoC improves the alignment between USTokens and text tokens in the shared latent space, fostering stronger synergy between understanding and generation representations.
% This approach not only enhances adaptability during dynamic training by exposing the model to diverse conditioning scenarios but also ensures effective alignment between DualToken and text token representations in the shared latent space, helping the model learn synergistic relationships between them.
The acoustic loss in AcousticGPT is defined as:
\begin{equation}\label{eq_llm_acousticloss}
\begin{aligned}
\mathcal{L}_{\text{acoustic}} = -\sum_{t=1}^{L} \log p(\mathbf{A^{tar}}_t | \mathbf{A^{tar}}_{<t}, \mathbf{S}, \mathbf{Spk}; \theta),
\end{aligned}
\end{equation}
where $\mathbf{S} \in \{\mathbf{S}_p, \mathbf{U^{tar}}, [\mathbf{S}_p; \mathbf{U^{tar}}]\}$ is selected via CoC, $\mathbf{Spk}$ is the speaker embedding, and $\mathbf{A^{tar}}_t$ is the predicted acoustic token at time $t$.
$p(\mathbf{A^{tar}}_t | \mathbf{A^{tar}}_{<t}, \mathbf{S}, \mathbf{Spk}; \theta)$ is the predicted probability of generating $\mathbf{A^{tar}}_t$, given $\{\mathbf{A^{tar}}_{<t}, \mathbf{S}, \mathbf{Spk}\}$.
Finally, the overall generation loss combines semantic and acoustic objectives as:
\begin{equation}\label{eq_llm_gen}
\begin{aligned}
\mathcal{L}_{\text{generation}} = 
\lambda \cdot \mathcal{L}_{\text{semantic}} + 
\xi \cdot \mathcal{L}_{\text{acoustic}},
\end{aligned}
\end{equation}
where $\lambda, \xi$ are weighting hyperparameters.

\section{Experiment Setup}
\subsection{Dataset}
% \subsubsection{Data for USTokenizer}

% During USTokenizer training, we used a dataset comprising nearly 2,000 hours of speech, jointly optimizing multiple speech understanding tasks, including Automatic Speech Recognition (ASR), Speech Emotion Recognition (SER), and Speech Question Answering (SQA). 
% Specifically, the LibriSpeech~\cite{7178964} train set was used for ASR training, while the IEMOCAP~\cite{busso2008iemocap} Sessions 1–4 were employed for SER. 
% The SQA task was trained on the SQA dataset from SALMONN~\cite{tang2023salmonn}, where questions were automatically generated from the textual content of LibriSpeech using ChatGPT. 
% In the SQA task, the model receives spoken content and a textual question as input and produces a natural language answer in text form.
% % In this setting, the LLM in USTokenizer is required to generate text responses to prompts by jointly considering both the speech input and the corresponding question.

We train the USTokenizer using multiple speech understanding tasks, including Automatic Speech Recognition (ASR), Speech Emotion Recognition (SER), and Speech Question Answering (SQA). They are based on LibriSpeech~\cite{7178964}, IEMOCAP~\cite{busso2008iemocap} and SQA dataset from SALMONN~\cite{tang2023salmonn}, respectively.
In the SQA dataset, both questions and answers were automatically generated based on LibriSpeech transcripts using ChatGPT, forming spoken-question–text-answer pairs.

DualSpeechLM is trained on eight speech understanding and generation tasks using approximately 4,500 hours of speech data. 
For understanding, we use LibriSpeech for ASR task, the English-to-German (En2De) subset of CoVoST2~\cite{wang2020covost} for Speech-to-Text Translation (S2TT) task, IEMOCAP~\cite{busso2008iemocap} 
% (Sessions 1–4 for training, Session 5 for evaluation)
for the SER task, and the SQA dataset~\cite{tang2023salmonn} constructed from LibriSpeech using ChatGPT. 
For generation, we train Text-to-Speech (TTS) on LibriTTS-R~\cite{koizumi2023librittsr}, Text-to-Speech Translation (T2ST) on Spanish-to-English (Es–En) and French-to-English (Fr–En) subsets of CVSS~\cite{jia2022cvss}, and Voice Conversion (VC) on a 1,000-hour subset of LibriHeavy~\cite{kang2023libriheavy} with evaluation on 400 pairs from VCTK~\cite{Yamagishi2019CSTRVC}. 
Additionally, the Speech Conversation (SC) task is built by synthesizing speech from the SQA dataset using the Volcengine TTS API. 
% Further details of the datasets are provided in the \textit{Appendix C}.
Please see \textit{Appendix C} for dataset details.

\subsection{Evaluation Metrics}

\subsubsection{Objective Evaluation}
For evaluation of the understanding tasks, we use Word Error Rate (WER) for ASR, BLEU-4 for S2TT, accuracy (ACC) for SER, and both BLEU-4 and ChatGPTScore for SQA, respectively.
For the generation tasks, we employ speaker similarity (SIM), WER, and DNSMOS to assess TTS and VC performance. 
BLEU-4 is used for T2ST. Both BLEU-4 and ChatGPTScore are used for evaluating the SC task.

\subsubsection{Subjective Evaluation}
We also conduct subjective evaluations for generation tasks, focusing primarily on TTS and VC. 
For each task, subjective assessment is carried out from speech quality (QMOS) and speaker similarity (SMOS).
More details can be found in \textit{Appendix C}.

\subsection{Model Settings}
% In the framework of DualSpeechLM, we conduct experiments using HuBERT tokens and our proposed USTokens as inputs, with acoustic tokens extracted by WavTokenizer as the output targets, respectively.
% In the following experiments, we refer to these two experiments as DualSpeechLM-Hubert and DualSpeechLM-USToken, respectively.
% Additionally, we trained a baseline under the same experimental conditions, where both the input and output of the text LLM are acoustic tokens of WavTokenizer, while maintaining all other configurations identical. 
In experiments, we compare four variants under the same configurations: (1) Baseline-Acoustic, where both input and output are acoustic token from WavTokenizer~\cite{ji2024wavtokenizer}; 
(2) Baseline-Semantic, where both input and output are HuBERT token~\cite{hsu2021hubert}; 
(3) DualSpeechLM-Hubert, using HuBERT token as input, acoustic token from WavTokenizer as output; and (4) DualSpeechLM-USToken, using our USToken as input, acoustic token from WavTokenizer as output.
% We adopt WavTokenizer~\cite{ji2024wavtokenizer}, 
% % a state-of-the-art discrete codec, 
% as the acoustic codec for AcousticGPT. 
% % The predicted acoustic tokens can then be directly decoded into speech signals via WavTokenizer.
Notably, the Baseline-Acoustic model requires 160k steps to converge under multi-task settings, while others converge in 60k steps.
Further details are provided in the \textit{Appendix D}.

\section{Results and Analyses}
\subsection{Understanding and Generation Performance}

\begin{table}[tb]
\small
\centering
\setlength{\tabcolsep}{0.3mm}
\begin{tabular}{cccccccc}
\toprule
\multirow{3}{*}{\textbf{Model}}            & \multirow{3}{*}{\textbf{LLM}}                                        & \multirow{3}{*}{\textbf{Token}}          & \multicolumn{2}{c}{\textbf{ASR}}                                       & \textbf{S2TT}                                      & \textbf{SER}               & \textbf{SQA}         \\ \cline{4-8} 
                                  &                         &                         & Clean             & Other             & \multicolumn{1}{c}{En2De}                  & -                 & -

\\ 
                                  % \cline{4-9} 
                                  &                          &                           & \multicolumn{2}{c}{$wer\downarrow$}                       & $bleu4\uparrow$                                  & $acc\uparrow$                & $bleu4\uparrow$    
                                  
                                  \\ \midrule

\multirow{2}{*}{\textbf{Ours}}            & \multicolumn{1}{c}{\multirow{2}{*}{Phi3.5-3B}} & Hubert                                              & \multicolumn{1}{c}{5.56}                   & 14.62                  & \multicolumn{1}{c}{10.20}                                        & 51.77                   & 42.59                   \\  
                                  
                                  & \multicolumn{1}{c}{}                                       & USToken                                                                  & \multicolumn{1}{c}{\underline{4.22}} & \underline{9.71}                   & \multicolumn{1}{c}{\underline{19.74}} &  \textbf{60.92}                   & \underline{44.38} \\ 
                                  \hline
\multirow{2}{*}{\textbf{Ours}}            & \multirow{2}{*}{Vicuna-7B}                                  & Hubert                                             & \multicolumn{1}{c}{5.28}                   & 16.64                  & \multicolumn{1}{c}{10.92}                                & 54.08                   & 42.68                   \\  
                                  &                                                             & USToken                                            & \multicolumn{1}{c}{\textbf{4.15}} & \textbf{9.69}                   & \multicolumn{1}{c}{\textbf{20.46}} & \underline{58.42}                   & \textbf{44.60}          \\ 
                                  \bottomrule
\end{tabular}
\caption{The understanding performance comparison of DualSpeechLM when using different text LLM backbones.}
\label{main_for_understanding_textllm}
\vspace{-5pt}
\end{table}

\begin{table*}[tb]
\small
\centering
\begin{tabular}{ccccccccc}
\toprule
      \multirow{3}{*}{\textbf{Model}}             &    \multirow{3}{*}{\textbf{LLM}}            &  \multirow{3}{*}{\textbf{Input Token}}                                                & \multicolumn{2}{c}{\textbf{TTS}}               & \textbf{VC}              & \multicolumn{2}{c}{\textbf{T2ST}}                                                          & \textbf{SC}        \\ \cline{4-9} 
           &                          &                           & \multicolumn{1}{c}{Clean}           & Other           & VCTK            & \multicolumn{1}{c}{Es2En} & Fr2En & - \\ 

% \cline{4-9} 
                         &  &     & \multicolumn{2}{c}{$s\uparrow/w\downarrow/d\uparrow$}
                                  & $s\uparrow/w\downarrow/d\uparrow$
                                  & \multicolumn{2}{c}{$bleu4\uparrow$}
                                  & $bleu4\uparrow$
                                  \\
           
           \midrule
           \multirow{2}{*}{\textbf{DualSpeechLM}}                                & \multirow{2}{*}{Phi3.5-3B}                          & Hubert  
                                & \multicolumn{1}{c}{0.81/20.80/3.35} & 0.81/17.12/3.38 & \textbf{0.82}/18.70/3.37 & \multicolumn{1}{c}{17.54}                         & 15.97                         & 11.06        \\  
                                 
                                  & \multicolumn{1}{c}{}                          &                   USToken                     & \multicolumn{1}{c}{\textbf{0.90}/\textbf{9.25}/3.86}  & \textbf{0.88/9.88}/\underline{3.82}  & \underline{0.80}/\textbf{10.16}/\textbf{3.46} & \multicolumn{1}{c}{\underline{26.77}}                         & \underline{23.82}                         & \underline{16.24}       \\ \hline
\multirow{2}{*}{\textbf{DualSpeechLM}}           
& \multirow{2}{*}{Vicuna-7B}                        
& Hubert                                               & \multicolumn{1}{c}{0.87/13.58/\underline{3.88}} & 0.86/15.56/3.80 & 0.78/19.86/3.38 & \multicolumn{1}{c}{22.29}                         & 19.74                         & 13.17         \\ 
                                  &                                                &   USToken                           & \multicolumn{1}{c}{\underline{0.89}/\underline{12.63}/\textbf{3.90}} & \underline{0.87}/\underline{14.65}/\textbf{3.86} & 0.78/\underline{17.03}/\underline{3.43} & \multicolumn{1}{c}{\textbf{29.58}}                         & \textbf{25.40}                         & \textbf{17.18}      \\ \bottomrule
\end{tabular}
\caption{Comparison of generation performance of DualSpeechLM using different text LLM backbones.}
\label{main_for_generation_textllm}
\end{table*}

\begin{table*}[tb]
\small
\centering
\setlength{\tabcolsep}{0.7mm}
\begin{tabular}{ccccccc}
\toprule
      \multirow{3}{*}{\textbf{Model}}                        & \multicolumn{2}{c}{\textbf{TTS}}               & \textbf{VC}              & \multicolumn{2}{c}{\textbf{T2ST}}                                                          & \textbf{SC}        \\ \cline{2-7} 
           &                    \multicolumn{1}{c}{Clean}           & Other           & VCTK            & \multicolumn{1}{c}{Es2En} & Fr2En & - \\ 

% \cline{4-9} 
    & \multicolumn{2}{c}{$s\uparrow/w\downarrow/d\uparrow$}
                                  & $s\uparrow/w\downarrow/d\uparrow$
                                  & \multicolumn{2}{c}{$bleu4\uparrow$}
                                  & $bleu4\uparrow$
                                  \\
           
           \midrule
           \textbf{DualSpeechLM}                                                                          & \multicolumn{1}{c}{0.90/9.25/3.86}  & 0.88/9.88/3.82  & 0.80/10.16/3.46 & \multicolumn{1}{c}{26.77}                         & 23.82                         & 16.24       \\ \hline

\textbf{USTokenizer Ablation} & &&&&& \\

w/o Understanding-driven Loss        
& 0.90/8.59/3.86 & 0.88/9.45/3.81 & 0.82/9.90/3.39 & 24.03 & 22.73 & 15.59
\\  
w/o Reconstruction Loss            &  0.83/52.50/3.02 & 0.81/54.24/2.96 & 0.79/26.08/3.33 & 32.43 & 29.46 &  17.09                                                           
\\

\hline

           \textbf{DualSpeechLM Ablation} & &&&&& \\

w/o Semantic Loss              &
0.80/167.56/3.85  & 0.85/175.11/3.83 & 0.80/264.53/3.45 & 0.09 & 0.07 & 0.15
\\ 
           
           w/o CoC              & 0.89/9.96/3.84 & 0.87/10.34/3.81 & 0.79/13.06/3.38 & - & - &- \\

           \bottomrule
\end{tabular}
\caption{Ablation study using generation tasks. We use Phi-3.5-3B as the text LLM backbone.}
\label{ablation_for_generation}
\vspace{-5pt}
\end{table*}

\begin{table}[tb]
\small
\centering
\setlength{\tabcolsep}{0.2mm}
\begin{tabular}{cccccc}
\toprule
\multirow{3}{*}{\textbf{Model}}            
% & \multirow{3}{*}{\textbf{LLM}}    
& \multicolumn{2}{c}{\textbf{ASR}}                                       & \textbf{S2TT}                                      & \textbf{SER}               & \textbf{SQA}         \\ \cline{2-6}                         & Clean             & Other             & \multicolumn{1}{c}{En2De}                                     & -                 & -

\\                          & \multicolumn{2}{c}{$wer\downarrow$}                       & $bleu4\uparrow$                                  & $acc\uparrow$                & $bleu4\uparrow$    
                                  
                                  \\ \midrule

\textbf{DualSpeechLM}            &  \multicolumn{1}{c}{4.22} & 9.71                   & \multicolumn{1}{c}{19.74}  & 60.92                   & 44.38 \\ \hline

\textbf{USTokenizer Ablation} & &&&& \\

w/o Understanding-driven Loss                                                                                                  & 4.81 & 10.43                   & 14.81 & 52.7                   & 37.67 \\ 

w/o Reconstruction Loss                                                                                                     & 4.74 & 10.46                   & 18.5 &  45.21                   & 46.44 \\ 

\textbf{DualSpeechLM Ablation} & &&&& \\

w/o Semantic Loss   & 4.31 & 9.61  & 18.67  & 60.35   & 43.86         \\ 
                                
                                  \bottomrule
\end{tabular}
\caption{Ablation study using understanding tasks. Phi-3.5-3B is used as the text LLM backbone.}
\label{ablation_for_understanding}
\vspace{-15pt}
\end{table}

We first compare the performance of our method with prior work. 
The results for understanding and generation are shown in Table \ref{main_for_understanding}, \ref{main_for_generation} and \ref{main_for_generation_subjective}, respectively.
From these results, we observe the following:

(1) \textit{USToken has better modality commonality with text, reducing the difficulty of modality alignment when adapting text LLMs to speech LLMs.}
DualSpeechLM-USToken outperforms both Baselines and DualSpeechLM-Hubert across almost all tasks, highlighting USToken's enhanced semantic understanding capabilities. 
Additionally, on translation tasks (S2TT and T2ST), which require the inherent translation capabilities of text LLM, our DualSpeechLM-USToken still significantly outperforms both Baselines and DualSpeechLM-Hubert. 
This demonstrates that USToken has better alignment with text modality, allowing it to retain more capabilities of text LLM during training. 
Overall, both DualSpeechLM-Hubert and DualSpeechLM-USToken exhibit better performance than the baseline, indicating that the design of Acoustic GPT in DualSpeechLM also helps alleviate the pressure of text LLM.
The comparison of convergence speed is provided in \textit{Appendix A}.
% In generation tasks, the performance advantage of DualSpeechLM-USToken is even more obvious than DualSpeechLM-Hubert, which can be attributed to the reconstruction loss added during training in USTokenizer. This helps preserve fine-grained semantic information, enabling better recovery for higher-quality speech synthesis.

(2) \textit{DualSpeechLM achieves strong performance in both tasks with small-scale resources.} 
Specifically, compared to Baselines, both DualSpeechLM-Hubert and DualSpeechLM-USToken demonstrate significant performance improvements, highlighting the effectiveness of DualSpeechLM. 
Compared to existing unified models, which rely on larger datasets and full fine-tuning, such as SpeechGPT and Mini-Omni2, DualSpeechLM-USToken achieves better performance with just 4.5k hours of data using parameter-efficient fine-tuning, LoRA. 
% This demonstrates that our approach can achieve strong performance with minimal resources and data.

(3) \textit{DualSpeechLM simultaneously fulfill the distinct information requirements of generation and understanding by dual-token modeling.}
% As shown in Ours in Figure~\ref{fig:preliminary}, the overall performance trends for both the generation and understanding tasks improve as the data volume for either understanding or generation increases, indicating that our approach can simultaneously enhance performance across both tasks.
% Overall, these results in Table~\ref{main_for_understanding} and Table~\ref{main_for_generation}, when combined with Figure~\ref{fig:preliminary}, showcase that our proposed method enables mutual enhancement between understanding and generation tasks in a unified model.
Comparisons between Baseline-Acoustic and Baseline-Semantic in Table \ref{main_for_understanding}, \ref{main_for_generation}, and \ref{main_for_generation_subjective} show that semantic tokens excel in understanding tasks, while acoustic tokens perform better in generation, e.g., higher SIM and DNSMOS scores in TTS. 
However, acoustic tokens carry weaker semantics reflected by higher WER results and lower translation quality. 
% Consequently, the Baseline-Acoustic also underperforms on tasks such as S2TT and T2ST, which require inherent text understanding from the text LLM.
% These results provide strong evidence that our proposed DualSpeechLM enables mutual enhancement between understanding and generation within a unified model.
In contrast, our DualSpeechLM demonstrates superior performance on both understanding and generation tasks simultaneously due to its dual-token modeling.
Another evidence is shown in Figure~\ref{fig:preliminary}, where our method consistently improves performance on both generation and understanding tasks as the amount of either type of training data increases. 
This demonstrates its strong ability to support one task without compromising the other.
% This indicates that our approach achieves stable and even enhanced performance across both tasks, regardless of which data type is scaled up.

% \subsection{Performance Comparison Across LLM Backbones}
\subsection{Performance with Different LLM Backbones}
% The results are presented in Table~\ref{main_for_understanding_textllm} and Table~\ref{main_for_generation_textllm}, corresponding to understanding and generation tasks, respectively.
% DualSpeechLM-USToken consistently outperforms DualSpeechLM-Hubert across almost all tasks, regardless of the underlying LLM. This demonstrates that USToken not only provides better semantic alignment with text LLMs, but also generalizes well across different model backbones. In particular, improvements are observed in both high-level understanding tasks (e.g., SQA, SER) and low-level generation tasks (e.g., TTS, VC), indicating that USToken effectively captures both semantic and acoustic characteristics. Notably, on challenging generation tasks such as S2TT, T2ST, and SC, DualSpeechLM-USToken shows substantial performance gains, further validating its strength in bridging speech and language modalities.
% Notably, it yields significant gains in both high-level understanding tasks (e.g., SQA, SER) and low-level generation tasks (e.g., TTS, VC), suggesting that USToken captures both semantic and acoustic characteristics.
% The improvements are especially pronounced in challenging generation tasks like S2TT, T2ST, and SC, further validating USToken’s effectiveness in modal commonality with text.
To further assess the generalizability of DualSpeechLM and the effectiveness of USToken across different backbones, we compare models using various text LLMs as backbones. 
Table \ref{main_for_understanding_textllm} and \ref{main_for_generation_textllm} show that DualSpeechLM-USToken consistently outperforms DualSpeechLM-Hubert across almost all tasks, regardless of the underlying LLMs.
This demonstrates that USToken not only aligns more effectively with text LLMs but also transfers well across model architectures. 
Additionally, our DualSpeechLM achieves performance comparable to or better than existing methods, with different text LLMs as the backbone. 
This highlights the strong compatibility and robustness of DualSpeechLM across different LLM backbones, further demonstrating its effectiveness as a unified framework for both speech understanding and generation.

\subsection{Ablation Study}
We perform ablation studies to evaluate the contribution of each component, with results shown in Table~\ref{ablation_for_generation} and~\ref{ablation_for_understanding}.
%we perform ablation studies using consistent architecture, datasets, and hyperparameters. Results are shown in Table~\ref{ablation_for_generation} and~\ref{ablation_for_understanding}.

\subsubsection{The Influence of Understanding-driven Loss. }  
We ablate the understanding-driven loss in USTokenizer while keeping all other components unchanged.
As shown in Table~\ref{ablation_for_understanding}, removing this LLM-based objective leads to significant performance drops in understanding tasks (e.g., BLEU4 decreases from 44.38 to 37.67 on SQA), highlighting its importance for aligning USTokens more closely with the semantic space of text LLMs using the understanding-driven loss.
For T2ST and SC in Table~\ref{ablation_for_generation}, we also observe a performance drop after removing the understanding-driven loss. 
This can be attributed to these two tasks rely more heavily on the internal capabilities of the text LLM. 
The understanding-driven loss plays a critical role in enhancing modality commonality between USTokens and text tokens, thereby helping to preserve as much of the original ability of the text LLM as possible when adapting the text LLM to speech-related tasks.
% Interestingly, Table~\ref{ablation_for_generation} shows slight improvements in tasks like TTS and VC, suggesting that training with reconstruction loss helps preserve more fine-grained semantic information, which in turn benefits the recovery of detailed speech signals during generation.

\subsubsection{Effect of the Reconstruction Loss.}
% We observe a significant performance drop in SER, TTS, and VC, suggesting that the reconstruction loss helps retain fine-grained details beyond semantics, which are crucial for these tasks. 
% For instance, the WER in TTS and VC increases dramatically, indicating degraded speech fidelity.
% Interestingly, for tasks that heavily rely on the reasoning capabilities of the text LLM, such as T2ST, the performance improves when only the understanding-driven loss is retained. This may also be attributed to the fact that USTokenizer, trained solely with the understanding-driven loss, aligns better with the text modality.
% thereby reducing the modality gap between speech and text and allowing the text LLM to function more effectively.
The third row in Table~\ref{ablation_for_generation} and~\ref{ablation_for_understanding} shows the results when the reconstruction loss is removed from USTokenizer. 
For tasks like T2ST that rely more on the reasoning capabilities of the text LLM, performance actually improves. 
This suggests that removing the reconstruction objective pushes USTokenizer to better align with the text modality, thus narrowing the modality gap and allowing the LLM to operate more effectively on semantically rich tasks.
However, we observe notable performance degradation in SER, TTS, and VC tasks, indicating that the reconstruction loss is essential for preserving fine-grained information, which is crucial for these tasks.

\subsubsection{Effect of Semantic Loss.}
The fourth row in Table~\ref{ablation_for_generation} and~\ref{ablation_for_understanding} shows the results of removing semantic loss from DualSpeechLM. 
This leads to a slight decline in understanding performance, indicating that reconstructing USTokens, as enforced by semantic loss, provides useful semantic supervision that benefits understanding tasks.
More critically, we observe a substantial performance drop across all generation tasks. 
Without semantic loss, the text LLM struggles with producing high-quality USTokens, resulting in degraded inputs to AcousticGPT and ultimately poor generation of acoustic tokens.
These results underscore the pivotal role of semantic loss in ensuring accurate semantic representation, which is essential not only for understanding but also for maintaining high-fidelity generation in unified DualSpeechLM.

\subsubsection{Influence of CoC.}
The final row in Table~\ref{ablation_for_generation} shows the impact of removing the Chain of Condition (CoC) strategy from the AcousticGPT module, which is applied only to TTS and VC tasks. 
Performance drops notably on both tasks, indicating that CoC's stochastic conditioning improves alignment between DualTokens and text tokens in the shared latent space. 
This enhanced alignment leads to more stable and accurate acoustic token generation.

\section{Conclusions}
In this work, we aim to develop a unified speech LLM that can simultaneously excel in both speech understanding and generation.
% present DualSpeechLM, a unified framework for joint speech understanding and generation via a novel dual-token modeling strategy. 
We first propose 
% a speech tokenization method, 
USTokenizer, 
% an Understanding-driven Speech Tokenizer, 
which reduces the modality gap between speech and text when adapting text LLMs to speech LLMs.
Based on this, we present an end-to-end DualSpeechLM that effectively accommodates the different informational requirements for understanding and generation by a dual-token modeling strategy. 
Experiments indicate that our method achieves significantly better performance than baselines, enabling mutual improvements between understanding and generation.
% achieving superior performance with significantly fewer resources. 
In the future, we plan to improve DualSpeechLM to larger, more diverse multilingual and cross-domain datasets to further explore its generalization and robustness.
% more expressive speech generation, more controllable

% \input{section/acknowledge}
\bibliography{aaai2026}

\clearpage
\newpage
\appendix
\section{Appendix}

\section{A. Convergence Speed Comparison}
\begin{figure}[htb]
\centering
\includegraphics[scale=0.5]
{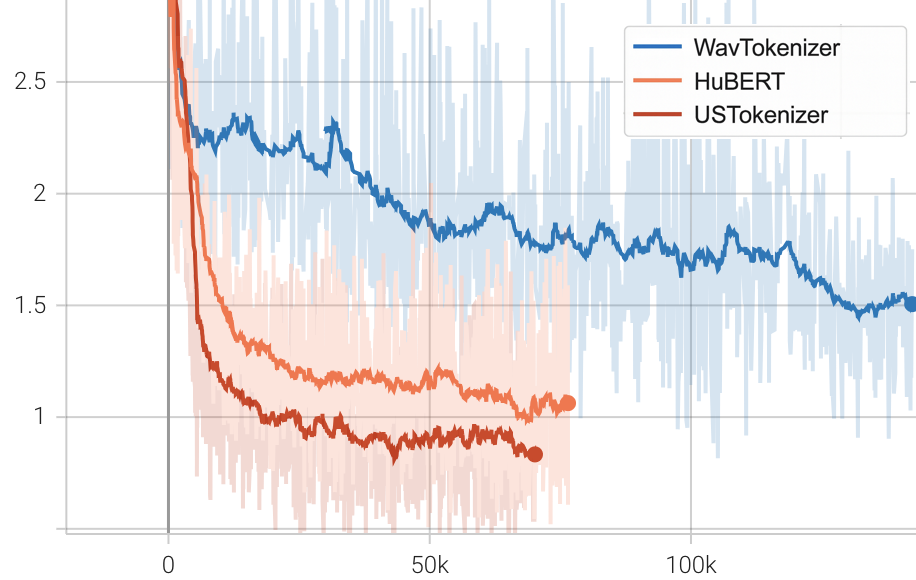}
\caption{Training loss curves for the understanding task. From top to bottom, the curves correspond to models trained with WavTokenizer, HuBERT, and USTokenizer as input, respectively.}
\label{fig:training_loss}
\end{figure}

Figure~\ref{fig:training_loss} shows the training loss curves of adapting text LLMs to speech LLMs with different tokens as input. The speech LLMs with USTokenizer and HuBERT converge faster and achieve lower final loss than other Speech LLMs using WavTokenizer. 
Notably, USTokenizer achieves the fastest convergence at the lowest loss, underscoring its effectiveness in adapting text LLMs to speech LLMs. 
This advantage stems from USTokenizer’s alignment with the semantic space of text LLMs, enabling USTokens to share stronger modality commonality with text and thus facilitating more efficient learning.

\section{B. Model Details}
\subsection{Vector Quantization}
In this work, we use a single quantizer. 
As shown in the architecture of USTokenizer in our main paper, the input $\mathbf{H}$ is first reshaped and passed through the codebook, where the distance between the input and the codebook vectors is computed. 
The nearest codebook vector is selected for each input element, producing the quantized output $\mathbf{Z}$. 
The commitment loss in VQ modules encourages the encoder's continuous outputs to stay close to the codebook vectors, stabilizing training. Based on this, we define the commitment loss as:
\begin{equation}\label{eq_tokenizer_commit}
\begin{aligned}
\mathcal{L}_{\text{commit}} = \left\| \mathbf{H} - \text{sg}(\mathbf{Z}) \right\|_2^2,
\end{aligned}
\end{equation}
where $\mathbf{H}$ is the continuous feature vector from the encoder,
$\mathbf{Z = vq(H)}$ is the quantized vector from the codebook,
$\text{sg}(\cdot)$ represents stop-gradient operator, that treats $\mathbf{Z}$ as a constant during backpropagation,
and $\|\cdot\|_2$ is the L2-norm.
By this commitment loss, the encoder parameters are updated to move $\mathbf{H}$ closer to $\mathbf{Z}$ during training.

\subsection{AcousticGPT}
\begin{figure}[htb]
\centering
\includegraphics[scale=1.0]
{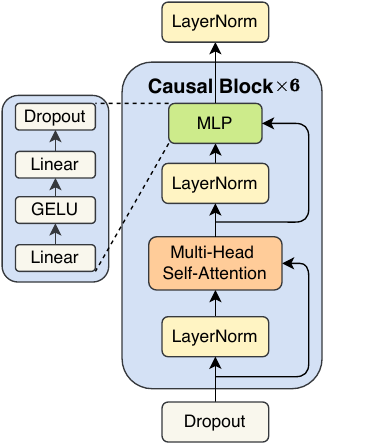}
\caption{The details of AcousticGPT .}
\label{fig:acousticgpt}
\end{figure}
The AcousticGPT module is implemented as a GPT-style structure, which autoregressively predicts next acoustic tokens based on previous acoustic tokens and conditioned on semantic representations and speaker embeddings. 
% In AcousticGPT, we also incorporate speaker embeddings extracted by 3D-Speaker~\cite{chen20243d} as a condition for autoregressive prediction.
As shown in Figure~\ref{fig:acousticgpt}, the module consists of six stacked causal blocks, each comprising a causal multi-head self-attention layer followed by a multi-layer perceptron (MLP).
Within each block, layer normalization is applied before both the self-attention and MLP modules, and residual connections are added after each to facilitate gradient flow and improve training stability. 
The MLP component includes a linear layer, GELU activation, another linear projection, and dropout for regularization. 
A final layer normalization layer is applied after the last causal block. 
During training, the hidden states extracted from text LLMs are first selected using the CoC strategy and then concatenated with speaker embeddings extracted by a pre-trained 3D-Speaker encoder~\cite{chen20243d} to form the input to the AcousticGPT module.
Then the model generates acoustic tokens that are coherent in both content and speaker characteristics. 
This architecture enables AcousticGPT to produce high-fidelity speech representations aligned with semantic and stylistic information.
\section{C. DataSet Statistics}
\subsection{Training Data Statistics}

\begin{table*}[tb]
\centering
\setlength{\tabcolsep}{1.0mm}
\begin{tabular}{ccccc}
\toprule
\multirow{1}{*}{\textbf{Task}}            &
\multirow{1}{*}{\textbf{Description}}            &
\multirow{1}{*}{\textbf{Data Source}}                       & \textbf{\#Hours}    & \textbf{\#Samples}                            

\\ 
                            \midrule

% \textbf{Pre-training Stage} \\
% Speech-Text & - & Libriheavy~\cite{kang2023libriheavy} &  40K & 10M \\
% \hline

% \textbf{Training Stage} \\
% \hdashline
\textbf{Understanding} \\
ASR & Automatic Speech Recognition & Librispeech~\cite{7178964}& 960 & 280K \\

S2TT &Speech-to-Text Translation & CoVoST2-En2De~\cite{wang2020covost} & 430 & 290K \\
SER & Speech Emotion Recognition &  IEMOCAP~\cite{busso2008iemocap} Session 1$\sim$4 & 5 & 4K\\
SQA & Speech Question Answering  & 
% Librispeech
SQA~\cite{tang2023salmonn}
& 960 & 280K \\

\textbf{Generation} \\

TTS & Text-to-Speech & LibriTTS-R~\cite{koizumi2023librittsr} & 960 & 350K\\

\multirow{2}{*}{T2ST} & \multirow{2}{*}{Text-to-Speech Translation} &  CVSS~\cite{jia2022cvss}-Es2En
& 60 & 70K\\

 &  &  CVSS~\cite{jia2022cvss}-Fr2En
& 109 & 130K\\

VC & Voice Conversion & Libriheavy~\cite{kang2023libriheavy} & 1K & 290K\\
SC & Speech Conversation & Synthetic SQA~\cite{tang2023salmonn} & 100 & 28K\\ 
\hdashline
\textbf{Total} &&& 4.5K & 1.7M \\
\bottomrule
\end{tabular}
\caption{Training data statistics. }
\label{exp_data-statistic1}
\end{table*}

\begin{table*}[tb]
\centering
\setlength{\tabcolsep}{1.0mm}
\begin{tabular}{cccc}
\toprule
\multirow{1}{*}{\textbf{Task}}            &
\multirow{1}{*}{\textbf{Description}}            &
\multirow{1}{*}{\textbf{Data Source}}                      & \multicolumn{1}{c}{\textbf{Eval Metrics }}                                     

\\ 
                            \midrule

\textbf{Understanding} \\
\multirow{2}{*}{ASR} & \multirow{2}{*}{Automatic Speech Recognition} & Librispeech test clean& \multirow{2}{*}{WER} \\

 &  & Librispeech test other&  \\

S2TT &Speech-to-Text Translation & CoVoST2-En2De & BLEU4 \\
SER & Speech Emotion Recognition &  IEMOCAP Session 5 & ACC \\
SQA & Speech Question Answering  & 
% Librispeech test 
SQA~\cite{tang2023salmonn}
& BLEU4/ChatGPTScore \\
\textbf{Generation} \\

\multirow{2}{*}{TTS} & \multirow{2}{*}{Text-to-Speech} & LibriTTS-R test clean & \multirow{2}{*}{SIM/WER/DNSMOS}\\
 &  & LibriTTS-R test other & \\

\multirow{2}{*}{T2ST} & \multirow{2}{*}{Text-to-Speech Translation} &  CVSS-Es2En
& \multirow{2}{*}{BLEU4} \\
 &  &  CVSS-Fr2En
& \\

VC & Voice Conversion & VCTK~\cite{Yamagishi2019CSTRVC} & SIM/WER/DNSMOS\\

SC & Speech Conversation & Synthetic SQA~\cite{tang2023salmonn} &BLEU4/ChatGPTScore \\  

\bottomrule
\end{tabular}
\caption{Test data statistics for DualSpeechLM. }
\label{exp_data-statistic2}
\end{table*}

During the training of USTokenizer, we utilize a dataset comprising nearly 2,000 hours of speech, jointly optimizing for multiple spoken language understanding tasks, including Automatic Speech Recognition (ASR), Speech Emotion Recognition (SER), and Speech Question Answering (SQA). 
Specifically, the LibriSpeech~\cite{7178964} train set was used for ASR, while the IEMOCAP~\cite{busso2008iemocap} Sessions 1–4 are employed for SER. 
For the SQA task, we utilize a dataset from SALMONN~\cite{tang2023salmonn}, where questions and answers are automatically generated from the LibriSpeech transcripts using ChatGPT. 
In this setting, the LLM is required to generate text responses to prompts by jointly considering both the speech input and the corresponding text question. 
We refer to this task as Speech Question Answering (SQA).

As shown in Table~\ref{exp_data-statistic1}, in DualSpeechLM, we jointly train a unified model for both speech understanding and generation across eight distinct tasks, leveraging approximately 4,500 hours of speech data. 
For speech understanding, we use the LibriSpeech~\cite{7178964} dataset to train and evaluate the Automatic Speech Recognition (ASR) task; 
the English-to-German (En2De) split of CoVoST2~\cite{wang2020covost} for training and evaluating the Speech-to-Text Translation (S2TT) task; 
IEMOCAP~\cite{busso2008iemocap} Sessions 1–4 for training and Session 5 for evaluating the Speech Emotion Recognition (SER) task; 
Similarly, in the SQA task, our DualSpeechLLM is required to generate text responses to prompts by jointly considering both the speech input and the corresponding text question. 
We use the SQA dataset~\cite{tang2023salmonn}, constructed from LibriSpeech using ChatGPT, for both training and evaluation. 
Specifically, 400 data pairs are randomly selected from the dataset to form a separate test set, which does not overlap with the training set.
For speech generation, we employ the LibriTTS-R~\cite{koizumi2023librittsr} dataset for training and evaluating the Text-to-Speech (TTS) task; 
the Spanish-to-English (Es–En) and French-to-English (Fr–En) splits of ~\cite{jia2022cvss} for training and evaluating the Text-to-Speech Translation (T2ST) task. 
For the Voice Conversion (VC) task, we randomly select 1,000 hours from LibriHeavy~\cite{kang2023libriheavy} for training. 
To evaluate the VC task, we randomly construct 400 data pairs from VCTK~\cite{Yamagishi2019CSTRVC} by pairing utterances of the same sentence spoken by different speakers, as well as different utterances by the same speaker for speaker embeddings.
Furthermore, we randomly sample 100 hours of speech from the SQA training set and use the open-source Volcengine TTS API to synthesize both corresponding questions and answers into speech, forming speech-question–speech-answer pairs for the Speech Conversation (SC) task. For evaluation, we construct a separate SC test set by randomly selecting 400 additional data pairs and applying the same synthesis procedure.
In the SC task, the DualSpeechLLM is required to generate speech responses to speech prompts that include speech input and the corresponding speech question.

\subsection{Objective Evaluation Metrics}
\begin{table}[tb]
\centering
\begin{tabular}{ccc}
\toprule
\textbf{USTokenizer}  &
\textbf{Configuration}                                                
\\ 
                            \midrule
\textbf{Model Configuration} \\
Codebook size & 1024 \\
Codebook dimension & 1024 \\
VQ layers & 1 \\
Encoder dimension & 1024 \\
Decoder dimension & 1024 \\

\hdashline
\textbf{Optimization Configuration} \\

Global batch size & 32 \\
Optimizer & AdamW \\
Optimizer hyperparameter & $\beta_1=0.9, \beta_2=0.999$ \\
Warmup Steps &3000 \\
Peak learning rate & 3e-5  \\
Minimum learning rate & 1e-5 \\
Learning rate decay & 5e-2 \\
Numerical precision & bf16 \\
\bottomrule
\end{tabular}
\caption{USTokenizer Configuration.}
\label{config:tokenizer}
\end{table}

As shown in Table~\ref{exp_data-statistic2}, we adopt a set of objective metrics to evaluate both understanding and generation performance in DualSpeechLM.
Specifically, Word Error Rate (WER) is used in ASR, TTS, and VC to quantify transcription errors by calculating the proportion of substitutions, insertions, and deletions relative to the total number of words in the reference, where lower WER indicates higher transcription fidelity. 
For TTS and VC tasks, we first transcribe the generated speech using Whisper large-v3~\cite{radford2023robustwhisper} \footnote{https://huggingface.co/openai/whisper-large-v3} and then compute WER based on the transcribed output. 
BLEU4~\cite{papineni2002bleu} is employed in S2TT, T2ST, SQA, and SC to measure the n-gram overlap (up to 4-grams) between generated and reference texts, where higher scores reflect stronger textual alignment, especially useful for translation.
For SQA and SC, since the answers are typically generated based on the given speech input, BLEU4 can be used to evaluate the consistency between the generated and reference answers. However, as BLEU4 has limited coverage for diverse but correct expressions, we additionally adopt ChatGPTScore for further evaluation for SQA and SC tasks.
Specifically, we use ChatGPT 4.1~\cite{achiam2023gpt} to directly rate the relevance, fluency, and correctness of generated and reference answers, with the evaluation prompt detailed as shown in Figure \ref{lst:qa-prompt}.
For the AC task, we first use Whisper large-v3~\cite{radford2023robustwhisper} to transcribe speech answers into text, and then employ the same evaluation prompt with ChatGPT 4.1 to assess the results.
For the SER task, we use classification accuracy (ACC) to assess the proportion of correctly predicted emotion categories. 
For speaker similarity (SIM), used in TTS and VC, we employ the wavlm-base-plus-sv~\cite{chen2022wavlm} \footnote{https://huggingface.co/microsoft/wavlm-base-plus-sv} model to extract speaker embeddings from both the prompt and synthesized speech, and compute the cosine similarity between them to assess speaker similarity.
Finally, DNSMOS~\cite{dnsmos} is applied in TTS and VC to assess speech naturalness. 
It is a no-reference neural metric that predicts mean opinion scores (MOS) on a 5-point Likert scale, approximating human judgments of audio quality.

\subsection{Subjective Evaluation Metrics}
Considering the cost of human evaluation, we perform subjective assessments primarily for the TTS and VC tasks.
For each task, assessments are performed from two perspectives: speech naturalness/quality (QMOS) and speaker similarity (SMOS). 
(1)QMOS evaluates the speech’s overall quality, naturalness, and clarity, while disregarding differences in speaking voice and speaking style such as emotion and prosody.
(2)SMOS assesses how closely the speaker identity (i.e., timbre) in the generated speech matches the that in the reference speech, ignoring content, grammar, or audio fidelity of the generated speech.
The scoring follows a 5-point Likert scale (5: excellent, 4: good, 3: fair, 2: poor, 1: bad). 
For each task, 15 audio samples are randomly selected and rated by 20 human evaluators.

\begin{table}[tb]
\centering
\setlength{\tabcolsep}{0.5mm}
\begin{tabular}{ccc}
\toprule
\textbf{DualSpeechLM }            &
\textbf{Configuration}                               
\\ 
                            \midrule
    \textbf{Model Configuration} \\

Embedding dim in AcousticGPT & 1024 \\
Number of heads in AcousticGPT  & 8\\

Max context length (\#tokens) & 2000\\
LoRA rank & 16 \\
\hdashline
\textbf{Optimization Configuration} \\

Global batch size & 72 \\
Optimizer & AdamW\\
Optimizer hyperparameter & $\beta_1=0.9, \beta_2=0.95$ \\
Warmup Steps& 3500 \\
Peak learning rate & 1e-4 \\
Minimum learning rate & 1e-5 \\
Learning rate decay & 5e-2\\
Numerical precision & bf16 \\
\bottomrule
\end{tabular}
\caption{DualSpeechLM Configuration.}
\label{config-dualspeechlm}
\end{table}

\begin{figure*}[!t]
\centering
% \begin{minipage}{0.95\textwidth}
\begin{lstlisting}[numbers=none, breaklines=true, backgroundcolor=\color{gray!5}, frame=single]
You are a professional evaluator for Question Answering (QA) tasks.

Firstly, you will be given the following content:
- speech content (from which questions will be asked based on the speech content),
- questions,
- reference answers,
- and then I will give you some candidate answers generated by different models sequentially.

You need to evaluate the overall quality of every model's output across all samples.
Please assess the model's answers based on the following criteria:

(1) Relevance (1-100): How well do the answers address the questions?
(2) Fluency (1-100): How natural and grammatically correct are the answers?
(3) Correctness (1-100): How accurately do the answers reflect the information in the reference answers?

Compare and evaluate these candidate answers generated by different models. 
After reading through all the sample questions, please give a final score (1-100 
points) for each candidate's answers, taking all three criteria into account. You 
should also write a brief explanation for each score.
\end{lstlisting}
% \end{minipage}
\caption{Prompt for QA Evaluation using ChatGPTScore.}
\label{lst:qa-prompt}
\end{figure*}

\section{D. Model Configuration}
\subsection{USTokenizer Configuration}

In USTokenizer, the Whisper-medium encoder~\footnote{https://github.com/openai/whisper} first extracts 50-dimensional features. 
After applying a 2$\times$ downsampling, we encode one second of 16kHz audio into 25 frames. We set the codebook size to 1024, which results in a bitrate of 250 bps. We train the tokenizer using a frozen llama3.2-1B \cite{dubey2024llama} model for the understanding-driven loss. 
USTokenizer is trained on 4 NVIDIA A100-40GB GPUs for 500k steps with the configurations listed in Table~\ref{config:tokenizer}. 
For weighting hyperparameter in $\mathcal{L}_{\text{USTokenizer}}$, we set $\alpha$, $\beta$, and $\gamma$ as 1, 5 and 45 respectively during training.
Lower $\mathcal{L}_{\text{reconstruction}}$ degrades generation performance, while lower $\mathcal{L}_{\text{Under}}$ harms understanding, which is similar to the trends observed in Table~\ref{ablation_for_generation} and Table~\ref{ablation_for_understanding}.
Moreover, removing the speech emotion recognition (SER) task results in a 9\% drop in emotion accuracy within SpeechLM, indicating that task supervision has an influence on the encoded token information.
We adopt AdamW~\cite{loshchilov2017decoupled} ($\beta_1 = 0.9$, $\beta_2 = 0.999$) as the optimizer, with a peak learning rate of $3\mathrm{e}{-5}$, a minimum learning rate of $1\mathrm{e}{-5}$, weight decay of $0.05$, and 3,000 warmup steps at the start learning rate of $1\mathrm{e}{-6}$. 
After the warmup stage, the learning rate is scheduled with cosine decay.

\subsection{DualSpeechLM Configuration}
To verify the generalization of USToken and the effectiveness of the DualSpeechLM framework on different backbones, 
we conduct experiments using text LLMs of Vicuna-7B\footnote{https://huggingface.co/lmsys/vicuna-7b-v1.1} and Phi-3B\footnote{https://huggingface.co/microsoft/Phi-3.5-mini-instruct} \cite{abdin2024phi3technicalreporthighly}.
For AcousticGPT, we adopt 6 multi-head self-attention layers, each featuring an embedding dimension of 1024 and 8 attention heads. 
Causal masking is applied in each layer to effectively model sequential dependencies.
During training, we set the weighting hyperparameters both $\lambda$ and $\xi$ in $\mathcal{L}_{\text{generation}}$ to 1.
All DualSpeechLM models are trained on 4 NVIDIA A100-40GB GPUs for 60,000 steps. The detailed model and optimization configurations are summarized in Table~\ref{config-dualspeechlm}.
Furthermore, it is important to note that in SpeechGPT \cite{zhang2023speechgpt} (Table \ref{main_for_understanding} and Table \ref{main_for_generation}), certain tasks, such as S2TT and T2ST, are not inherently supported. 
During inference, the S2TT task is accomplished by first applying ASR to transcribe speech into text, followed by translating the resulting English text into German. 
For the T2ST task, we first translate the Spanish and French text inputs into English, and subsequently perform TTS to synthesize speech.

\section{E. Semantic Comparison of USTokenizer}

\begin{table*}[ht!]
\centering
\begin{tabular}{cc}
\toprule
\textbf{Task} & \textbf{Prompt} \\
\midrule
ASR & Recognize the speech and give me the transcription. \\
TTS & Please read this sentence out loud. \\
VC & Without altering the spoken content, transform the speaker's voice in this speech to match the target voice. \\
T2ST & Please translate the $[source\; language]$ text into $[target\; language]$ speech. \\
SC & Please listen to the speech content and provide a spoken answer to the question. \\
SQA & Based on the content, provide a text-based answer to the question. \\
S2TT & Please translate the $[source\; language]$ speech into $[target\; language]$ text transcription. \\
SER & Please describe the emotion of the speaker. \\
\bottomrule
\end{tabular}
\caption{Task prompts used in multi-task evaluation.}
\label{tab:task_prompts}
\end{table*}

\subsection{Evaluation with Recent Tokenizers}
To further validate the semantic capability of our USTokenizer, we retrain automatic speech recognition (ASR) models using several recent tokenizers on Librispeech, including TAAE~\cite{parkerscaling}, MimiCodec~\cite{defossez2024moshi}, and CosyVoice-2~\cite{du2024cosyvoice2}, under identical experimental setups.
For fairness, we only use the first quantization layer of MimiCodec, which is distilled from the first layer of WavLM~\cite{chen2022wavlm}.
As shown in Table \ref{res:asr_tokenizer_comparison}, USTokenizer achieves the lowest Word Error Rate (WER) on both clean and noisy speech conditions.
These results demonstrate that USTokenizer provides a more effective abstraction of high-level information, indicating its stronger ability in semantic preservation.

\begin{table}[tb]
\small
\centering
\begin{tabular}{ccc}
\toprule
\multirow{2}{*}{\textbf{Model}}  
                            & \textbf{Clean}             & \textbf{Other}         

\\ 
                                  % \cline{4-9} 
                         & \multicolumn{2}{c}{$wer\downarrow$}
                                  \\ \midrule

TAAE & 13.24 & 29.49          \\ 
MimiCodec &	5.96 & 14.54 \\
CosyVoice2 &	6.29&15.19 \\
\textbf{USTokenizer(Ours)} &	4.22&9.71 \\
                                  \bottomrule
\end{tabular}
\caption{The Comparison with Recent Tokenizers.}
\label{res:asr_tokenizer_comparison}
\end{table}

\subsection{Comparison with Baichuan-Audio Tokenizer}
We also compare USTokenizer with Baichuan-Audio~\cite{li2025baichuanaudio}, a recent unified framework for end-to-end speech interaction.
Baichuan-Audio adopts an 8-layer Residual Vector Quantization (RVQ) scheme optimized for Mel-spectrogram reconstruction.
In contrast, USTokenizer employs a single-layer VQ optimized for self-supervised representation reconstruction, which is simpler, more LLM-friendly, and emphasizes semantic preservation.
As summarized in Table~\ref{res:baichuan_comparison}, Baichuan-Audio exhibits a higher WER, possibly because Mel reconstruction prioritizes acoustic fidelity, which in turn diminishes semantic retention to some extent.
By contrast, USTokenizer reconstructs self-supervised representations, achieving a lower bitrate and richer semantic encoding.

\begin{table}[tb]
\small
\centering
\setlength{\tabcolsep}{0.6mm}
\begin{tabular}{ccccc}
\toprule
\multirow{2}{*}{\textbf{Model}}  & \textbf{Frame Rate}	& \textbf{Token}
& \textbf{Bitrate}	  & \textbf{Clean/Other}  \\
& $Hz$ & $num/s$ & $kbps$ & $wer\downarrow$

\\  \midrule

Baichuan-Audio & 12.5 &	100	& 1.08	& 4.53/10.65 \\
\textbf{USTokenizer(Ours)} & 25&	25	&0.25	&4.22/9.71 \\
                                  \bottomrule
\end{tabular}
\caption{The Comparison with Baichuan-Audio on  Librispeech.}
\label{res:baichuan_comparison}
\end{table}

\section{F. Computational Overhead Analysis}

To quantify the additional computational cost introduced by USTokenizer and DualSpeechLM, we report the relative changes in memory and training time under different configurations.
As shown in Table~\ref{tab:overhead}, the text-LLM module in USTokenizer introduces moderate overhead but remains manageable, as it is unused during inference and converges quickly when applied to DualSpeechLM.
Furthermore, DualSpeechLM demonstrates higher training efficiency owing to the low-bitrate token design of USTokenizer, which reduces both memory and time consumption compared with baselines.

\begin{table}[tb]
\small
\centering
\setlength{\tabcolsep}{0.6mm}
\begin{tabular}{cccc}
\toprule
\textbf{Model} & \textbf{Setting}
& \textbf{Memory(\%)}	  & \textbf{Time(\%)}  
\\  \midrule

USTokenizer & vs w/o Text LLM &$\uparrow$ 288	&$\uparrow$ 76  \\
DualSpeechLM & vs Baseline-Semantic & $\downarrow$ 26 &	$\downarrow$ 32
\\
DualSpeechLM & vs Baseline-Acoustic &
$\downarrow$ 17 &	$\downarrow$ 31
\\
                                  \bottomrule
\end{tabular}
\caption{Training overhead of USTokenizer and DualSpeechLM.}
\label{tab:overhead}
\end{table}

Overall, these results indicate that the additional text-LLM module in USTokenizer slightly increases training cost but does not affect inference efficiency, while DualSpeechLM achieves better computational efficiency.
\section{G. Prompts for different tasks}
Table~\ref{tab:task_prompts} summarizes the prompts used for different tasks in training and evaluation of our DualSpeechLM.
\section{H. Discussion}
Although Figure \ref{fig:preliminary} shows that understanding and generation tasks can mutually enhance each other in our framework, a closer look at subfigures (a) Right and (b) Left reveals an interesting asymmetry: increasing the data for understanding significantly boosts generation performance, whereas increasing generation data yields only limited improvements in understanding.
This may be because the understanding task is inherently `stronger' in our setting. When more understanding data is provided, the text LLM can better learn the mapping between text tokens and DualTokens, leading to more accurate DualToken representations. These improved semantic representations in turn enhance AcousticGPT’s predictions, thereby improving generation performance.
Conversely, even when generation data increases, its relatively weaker semantic quality limits the benefit it can bring to the understanding task. 
Nevertheless, our semantic supervision loss mitigates this issue by explicitly improving the model’s semantic representation capability during generation. This enables generation data to still have a positive, albeit smaller, effect on understanding.
In contrast, the baseline model in subfigure (b) Left relies solely on acoustic tokens, which lack sufficient semantic information, resulting in poor performance on understanding tasks.

\end{document}